\documentclass[manuscript,trackchanges]{aastex61}
\pdfoutput=1

\definecolor{DarkGreen}{rgb}{0.0,0.4,0.0}  

\newcommand{\ie}{{\it i.e.}~}
\newcommand{\eg}{{\it e.g.},~}

\newcommand{\bluec}[1]{{\color{blue}{#1}}}

\newcommand{\vect}[1]{\mathbf{#1}}

\hypersetup{
    colorlinks=true,       
     linkcolor=black,          
     citecolor=black,        
     filecolor=black,      
      urlcolor=black           
}

\received{September 12, 2018}
\revised{November 16, 2018}
\accepted{November 24, 2018}

\shorttitle{Evolution of an MFR}
\shortauthors{Wang et al.}

\begin{document}

\title{Evolution of a Magnetic Flux Rope toward Eruption}

\correspondingauthor{Wensi Wang}
\email{minesnow@mail.ustc.edu.cn}

\author{Wensi Wang}
\affiliation{CAS Key Laboratory of Geospace Environment, Department of Geophysics and Planetary Sciences\\
University of Science and Technology of China, Hefei, 230026, China}
\affiliation{Department of Physic, Montana State University, Bozeman, MT 59717, USA}

\author{Chunming Zhu}
\affiliation{Department of Physic, Montana State University, Bozeman, MT 59717, USA}

\author{Jiong Qiu}
\affiliation{Department of Physic, Montana State University, Bozeman, MT 59717, USA}

\author{Rui Liu}
\affiliation{CAS Key Laboratory of Geospace Environment, Department of Geophysics and Planetary Sciences\\
University of Science and Technology of China, Hefei, 230026, China}

\author{Kai E. Yang}
\affiliation{School of Astronomy and Space Science, Nanjing University, Nanjing, 210023, China}

\author{Qiang Hu}
\affiliation{Department of Space Science and CSPAR, University of Alabama in Huntsville, Huntsville, AL 35805, USA}

\begin{abstract}

It is well accepted that a magnetic flux rope (MFR) is a critical component of many coronal mass ejections (CMEs), yet how it evolves toward eruption remains unclear. Here we investigate the continuous evolution of a pre-existing MFR, which is rooted in strong photospheric magnetic fields and electric currents. The evolution of the MFR is observed by the Solar Terrestrial Relations Observatory (STEREO) and the Solar Dynamics Observatory (SDO) from multiple viewpoints. From STEREO's perspective, the MFR starts to rise slowly above the limb five hours before it erupts as a halo CME on 2012 June 14. In SDO observations, conjugate dimmings develop on the disk, simultaneously with the gradual expansion of the MFR, suggesting that the dimmings map the MFR's feet. The evolution comprises a two-stage gradual expansion followed by another stage of rapid acceleration/eruption. Quantitative measurements indicate that magnetic twist of the MFR increases from $1.0 \pm 0.5$ to $2.0 \pm 0.5$ turns during the five-hour expansion, and further increases to about 4.0 turns per AU when detected as a magnetic cloud at 1 AU two day later. In addition, each stage is preceded by flare(s), implying reconnection is actively involved in the evolution and eruption of the MFR. The implications of these measurements on the CME initiation mechanisms are discussed.

\end{abstract}


\section{Introduction} \label{sec:intro}

In-situ measurements of many coronal mass ejections (CMEs) near the Earth reveal a magnetic flux rope (MFR) structure, which possesses helical magnetic fields. These interplanetary CMEs are sometimes called magnetic clouds (MCs; \citealt{burlaga1981magnetic,burlaga1982magnetic}), a dominant contributor to adverse space weather when they are directed toward the Earth. Now it is generally accepted that MFRs originate from the Sun \citep{gosling1990coronal}. Therefore, it is of critical importance to understand how an MFR forms and evolves on the Sun, and what mechanism triggers its eruption. \par
 
Although it is believed that an MFR can form on the Sun, it is not clear when and how it forms. MFRs may emerge from the photosphere (\eg \citealt{fan2004numerical,fan2009emergence}), or form by magnetic reconnection (\eg \citealt{van1989formation}). They can stay in equilibrium in the corona until eruption, thus are called {\em pre-existing} MFRs. Observations in the past decades have provided evidences for pre-existing MFRs. Forward or reversed S-shaped sigmoids in soft X-rays or EUV images are sometimes observed hours before the eruption, and are often considered as a manifestation of the MFR topology \citep{rust1996evidence,aurass2000flares,green2007transient,liu2010sigmoid,green2011photospheric,savcheva2014new}. Filaments or prominences are thought to be plasma identities of MFRs \citep{vrvsnak1988structure,vrvsnak1991stability,low1995magnetostatic,dere1999lasco,liu2012slow}. Recent studies suggest that, for a prominence-cavity system, the cavity can be considered as a {\em pre-existing} MFR with prominence material embedded at the bottom of the flux rope (see the details in a review by \citealt{schmieder2015flare}). Lately, the so-called hot blob \citep{cheng2011observing}, which appears bright in high-temperature passbands ($T \geq 8$~MK) but is observed as a dark cavity in other low-temperature passbands, is also interpreted as an MFR (\eg \citealt{zhang2012observation,song2014temperature,nindos2015common}). Gradual development of pre-eruption dimmings is considered as another signature for the existence of a pre-existing MFR \citep{gopalswamy1999dynamical,qiu2017gradual}. \par 

The eruption of a pre-existing MFR may be triggered by an ideal MHD instability, such as torus instability \citep{bateman1978mhd, kliem2006torus} or kink instability (\eg \citealt{dungey1954twisted,hood1979kink,torok2005confined}). Torus instability may occur when the downward Lorentz force exerted upon the MFR by the envelope field $B_p$ decreases more steeply with height $h$ than the hoop force of the MFR current \citep{kliem2006torus}, \ie when the decay index ($n = - d \ln B_{p} / d \ln h $) exceeds a critical value $n_{crit}$. Analytical and numerical MHD models suggest this critical value falls in the range of 1 $-$ 2, depending on magnetic configurations \citep{fan2007onset,aulanier2010formation,demoulin2010criteria,zuccarello2016apparent}. For example, \cite{bateman1978mhd} and \cite{kliem2006torus} derived $n_{crit}$ of 1.5 for an idealized current ring. In MHD simulations, $n_{crit}$ is found to be in the range of 1.4 $-$ 2.0 \citep{fan2007onset,zuccarello2016apparent}. Also, an MFR may become kink-unstable when it acquires sufficient amount of twist, \ie the number of turns that the magnetic field lines wind around the rope axis. Theoretical and numerical studies have estimated the critical twist $T_{crit}$ in flux ropes with different magnetic configurations \citep{dungey1954twisted,hood1979kink,bennett1999waves,baty2001mhd,torok2004ideal,fan2005coronal,torok2005confined,kliem2010reconnection}. For example, \cite{hood1981critical} derived $T_{crit}$ of 1.25 turns in a nonlinear force-free MFR with the uniform-twist solution \citep{gold1960origin}. \cite{fan2004numerical} found that a line-tied flux tube which emerged into a coronal potential field erupted when its twist reached $1.76$ turns. \cite{torok2004ideal} provided $T_{crit} \approx 1.75$ turns in a force-free coronal loop model \citep{titov1999basic}. In a magnetostatic configuration ($\beta \neq 0$), the profile of pressure gradient and plasma compressibility also affect the stability \citep{newcomb1960hydromagnetic,hood1979kink,einaudi1983stability}. \par

The eruption of an MFR can also be triggered by magnetic reconnection. The main role of reconnection is to remove the field lines originally confining the MFR and thus to release it. Reconnection may occur below an MFR \citep{forbes1984numerical,forbes1990numerical,forbes1995photospheric,chen2000emerging,moore2001onset}, above an MFR (\eg \citealt{antiochos1999model}), or within a compound MFR system (\eg \citealt{liu2012slow,zhu2015complex,awasthi2018pre}). \par

On the other hand, a coherent MFR structure might not be present prior to its eruption, but is created by magnetic reconnection, \ie in-situ formed. For example, reconnection may occur between neighboring sets of field lines in a sheared magnetic arcade, resulting in a flux rope (\eg \citealt{mikic1988dynamical,mikic1994disruption,demoulin1996three,demoulin2002magnetic}). \cite{wang2017buildup} identified two newly-formed and expanding bright rings enclosing conjugate dimming regions, suggesting the dynamic formation of an MFR via reconnection. \cite{gou2018birth} observed that a leading plasmoid formed at the upper tip of a preexistent current sheet through the coalescence of plasmoids, and then expanded impulsively into a CME bubble. During the eruption, reconnection could add a significant amount of magnetic flux into an MFR \citep{lin2004role}, as implied by a statistical comparison between the flux budget of interplanetary magnetic clouds and the reconnection flux in their source regions \citep{qiu2007magnetic,hu2014structures}.\par

To help understand mechanisms for the formation and eruption of MFRs, it is important to measure the magnetic properties of MFRs. However, direct measurement of the coronal magnetic fields is still unavailable, thus it poses a major challenge to the observational investigation of the MFR properties. An alternative approach is to investigate the footpoints of MFRs which are anchored to the Sun (\eg \citealt{qiu2007magnetic,wang2017buildup}). In this study, combining remote-sensing observations and in-situ detection, we identify a pre-existing MFR, and quantify its magnetic and dynamic properties during five hours before its eruption. Both feet of the MFR are co-spatial with strong magnetic fields and electric currents. This unique event allows us to test the roles of reconnection and/or ideal MHD instabilities played in the evolution of the MFR. In the following text, Section~\ref{obs} highlights key observations. Section~\ref{method} describes the methods used to identify the feet of the MFR. Then we analyze quantitative measurements and evolution of the MFR in Section~\ref{ana}. Discussions and conclusions are given in Section~\ref{dis}. \par

\section{Overview of Observations} \label{obs}
\begin{figure}
\epsscale{.95}
\plotone{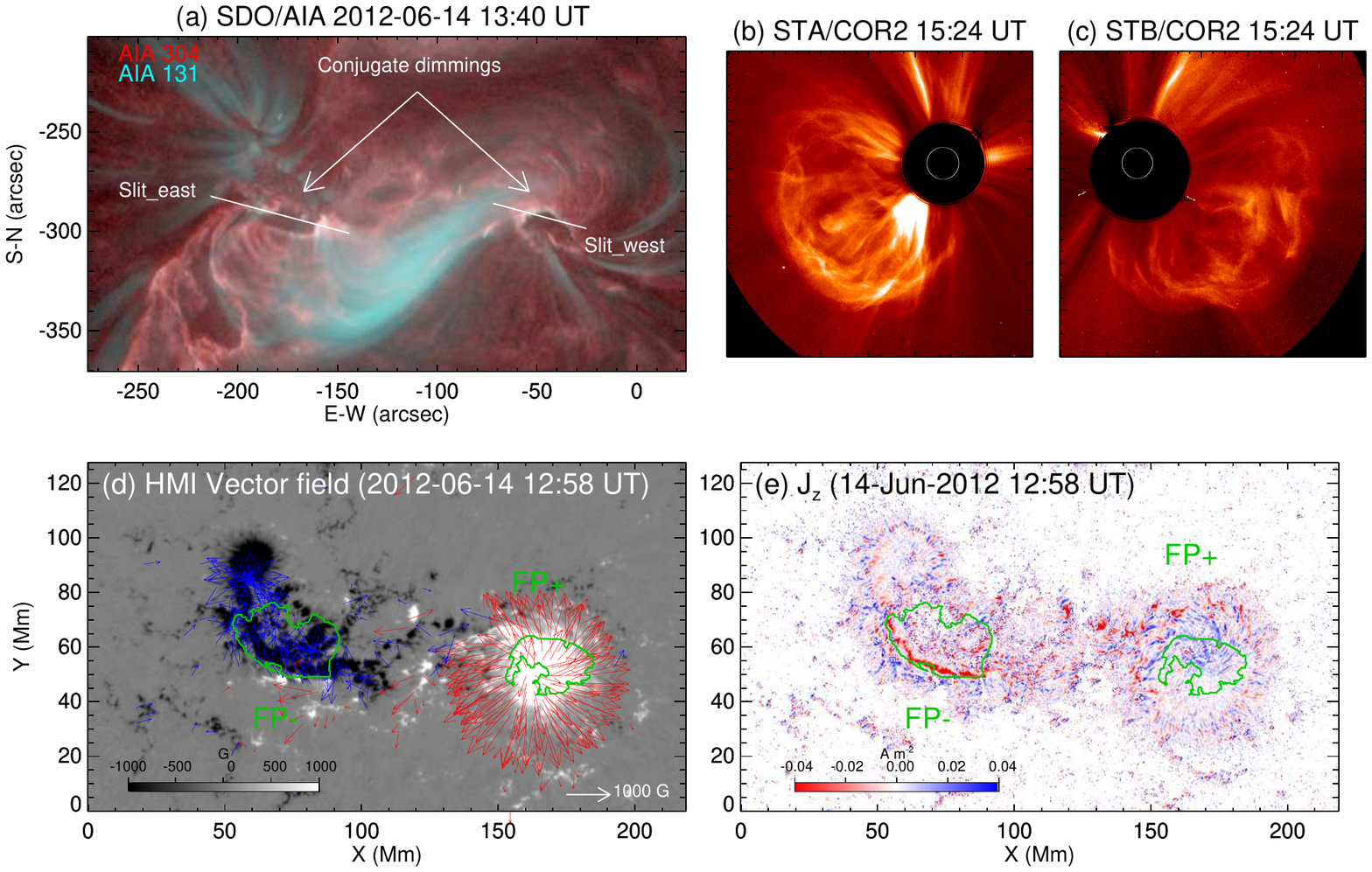}
\vskip -4cm
\caption{Overview of the event. (a) The M1.9 flare observed on 2012 June 14 in SDO/AIA 304 \AA\ (red) and 131 \AA\ (cyan). Two conjugate dimmings are labeled in (a). Two slits (white lines) are selected to construct the time-distance maps in Figure~\ref{fig3}. The M1.9 flare is associated with a halo CME, which was captured by STA/COR2 in (b) and STB/COR2 in (c) from different perspectives. (d) HMI vector magnetogram, with the transverse fields indicated with arrows. (e) The vertical current density (see details in Section~\ref{method}), with the positive (negative) values colored in blue (red). Two contours (green) in (d) or (e) outline the two conjugate feet of the MFR, labeled as `FP-' and `FP+'. \label{fig1}}
\end{figure}

The eruption of interest occurred in a sigmoidal active region (AR 11504) on 2012 June 14, which was observed by the Solar Dynamics Observatory (SDO; \citealt{pesnell2011solar}) and the Solar Terrestrial Relations Observatory (STEREO; \citealt{kaiser2008stereo}). The host active region is characterized by fast sunspot rotation with an average speed of 5 degrees hour$^{-1}$, exceeding the typical speed of 1 $-$ 2 degrees hour$^{-1}$ \citep{zhu2012velocity}, and shear flows lasting for a few days. It produced several confined flares and an eruptive M1.9 flare (Figure~\ref{fig1} (a)) on 2012 June 14. From SDO's perspective, the active region was located near the disk center (S17W00). The Atmospheric Imaging Assembly (AIA; \cite{lemen2012atmospheric}) onboard SDO takes full-disk images up to 1.5 $R_\odot$ at a spatial scale of $0''.6$ pixel$^{-1}$ in seven EUV channels with a cadence of 12 s and two UV channels with a cadence of 24 s spanning a broad range of temperature. In this paper, we mainly used 1600~{\AA} (C~{\sc iv} +  continuum; $\log T = 5.0,3.7 $), 304~{\AA} (He~{\sc ii}; $\log T = 4.7$) and 94~{\AA} (Fe~{\sc xviii}; $\log T = 6.9$) channels to study flare ribbons, coronal dimmings, and flare loops, respectively. The STEREO ``Ahead'' and ``Behind'' spacecraft (hereafter STA and STB) captured the eruption at its east and west limb on that day, respectively. Images taken by Extreme Ultraviolet Imager (EUVI; \cite{howard2008sun}) onboard STB in 195 passband with a 5-min cadence reveal a gradual expansion of a coronal structure (see Figure~\ref{fig2}) lasting for more than five hours before the eruption. The slowly expanding coronal structure finally evolved into a halo CME propagating at about 1000 km s$^{-1}$ when observed in white light by COR2 coronagraph onboard STEREO (Figures~\ref{fig1} (b) and (c)). The image from STA/COR2 (Figure~\ref{fig1} (b)) shows the CME with clear helical features. Two days later, it passed through WIND spacecraft at 1 AU  and is identified as an MC (\citealt{palmerio2017determining,james2017disc}), indicating that an MFR is involved in the eruption. \par

Coronal dimmings are observed in two regions with opposite magnetic polarities (Figure~\ref{fig1} (a)). The twin dimming regions (in 304~{\AA}) near the two ends of the S-shaped sigmoid (in 131~{\AA}) are labeled as ``conjugate dimmings''. To study their evolutions, time-distance maps along two slits (Figure~\ref{fig1} (a)) through the dimming regions are generated, as shown in Figure~\ref{fig3}. These maps demonstrate that the dimmings have started five hours before the eruption. The pre-eruption dimmings, as indicated by the normalized lightcurves of two dimming segments (Cut1 and Cut2 in Figure~\ref{fig3} (c1) and (d1), respectively), are clearly visible in almost all AIA channels (except 94~{\AA} in the western dimming region, see Figures~\ref{fig3} (a) and (b)), implying that the changes in the brightnesses are primarily due to a decrease in plasma density rather than a change in temperature. In addition, conjugate dimmings evolve simultaneously with the expanding coronal structure, suggesting that the dimmings map the feet of the coronal structure. It is noteworthy that the conjugate dimmings in this event are co-spatial with strong magnetic fields and vertical electric currents (Figures~\ref{fig1} (d) and (e)) measured with vector magnetograms obtained by the Helioseismic and Magnetic Imager (HMI; \citealt{scherrer2012helioseismic}). These observations  indicate that the coronal structure is associated with a pre-existing MFR, and allow us to quantify its magnetic properties and study its evolution in detail, as presented below. \par

\section{Identification of the MFR's feet}\label{method}
Similar to some previous studies \citep{webb2000relationship,qiu2007magnetic,cheng2016nature,wang2017buildup}, we utilize conjugate dimmings to identify the MFR's feet. Many studies suggest that observations in chromospheric/transition-region lines are less subject to coronal loops projection, compared with those in coronal lines \citep{harvey2002polar,qiu2007magnetic,harra2007coronal,scholl2008automatic}. We hence detect dimmings in the AIA 304 images at a cadence of 1 minute, starting from 07:00 to 18:00 UT. All AIA images are differentially rotated to the same moment (two hours before the M1.9 flare) using the standard SolarSoftware packages. \par

For each image, we select dimmed pixels within two regions at ends of the sigmoid by a thresholding method: a pixel is flagged when its brightness is reduced by 30\% compared with its original value averaged between 07:00 and 08:00 UT, and displays continuous decline until eruption. The two contours in Figures~\ref{fig1} (d) and (e) enclose all the identified dimming pixels. We thus consider the two contours ( `FP+' and `FP-') encompass maximum areas of the MFR's conjugate footpoints in the positive and negative fields, respectively. \par

The magnetic properties at the feet of the MFR are studied by projecting the identified dimming pixels onto HMI vector magnetograms. The magnetogram is disambiguated and deprojected to the heliographic coordinates with a Lambert (cylindrical equal area; CEA) projection method, resulting in a pixel scale of 0.36 Mm \citep{bobra2014helioseismic}. The magnetic fluxes ($\Phi_{z}$ in Table~\ref{tab:feet}) are calculated in two conjugate footpoints. In order to avoid the effect of noise in the HMI magnetogram, only the vertical magnetic fields with the strengths $|B_{z}| > 20$ G \citep{liu2012comparison,hoeksema2014helioseismic} are used. It is also noted that in each footpoint, the magnetic flux of one polarity over-dominates the other polarity by more than two orders of magnitude, therefore, each footpoint can be treated as monopolar. In addition, we calculate the vertical current density $J_{z}=(\bigtriangledown \times \vect{B})_{z} / \mu_{0}$ \ ($\mu_{0}=4\pi \times 10^{-7}$ H m$^{-1}$) (Figure~\ref{fig1} (e)) in the vector magnetogram mentioned before. All measurements are listed in Table~\ref{tab:feet} and will be discussed in detail in the next section. 


\section{Properties and Evolution of the MFR}\label{ana}
In the following sections, we investigate the evolution of the conjugate dimmings and the expanding coronal structure in detail, showing how the MFR evolve toward the eruption. Then we analyze the results of measurements and infer magnetic twist of the MFR during that period. In addition, we found obvious temporal and spatial relationships between flares and the evolution of the MFR. \par

\subsection{Dynamic evolution}
To show the dimming evolution, we plot the light-curve of the average brightnesses ($I_{FP+}$ and $I_{FP-}$) for the two fixed regions (`FP+' and `FP-' in Figure~\ref{fig1} (d)) respectively, as seen in Figure~\ref{fig4} (b). The dimmings go through three stages (marked in Figure~\ref{fig4} (b)). During the first stage, both $I_{FP+}$ and $I_{FP-}$  decrease simultaneously and gradually for about three hours. While in the second stage, $I_{FP+}$ and $I_{FP-}$ decrease more rapidly than in the previous one. About two hours later, $I_{FP+}$ reaches the minimum, which is $75 \%$ of its brightness at 08:00 UT. At 13:25 UT, $I_{FP-}$ reaches its minimum, which is $65 \%$ of its brightness at 08:00 UT. After that, both $I_{FP-}$ and $I_{FP+}$ start to increase, corresponding to the third stage. \par 

\begin{figure}
\epsscale{.9}
\plotone{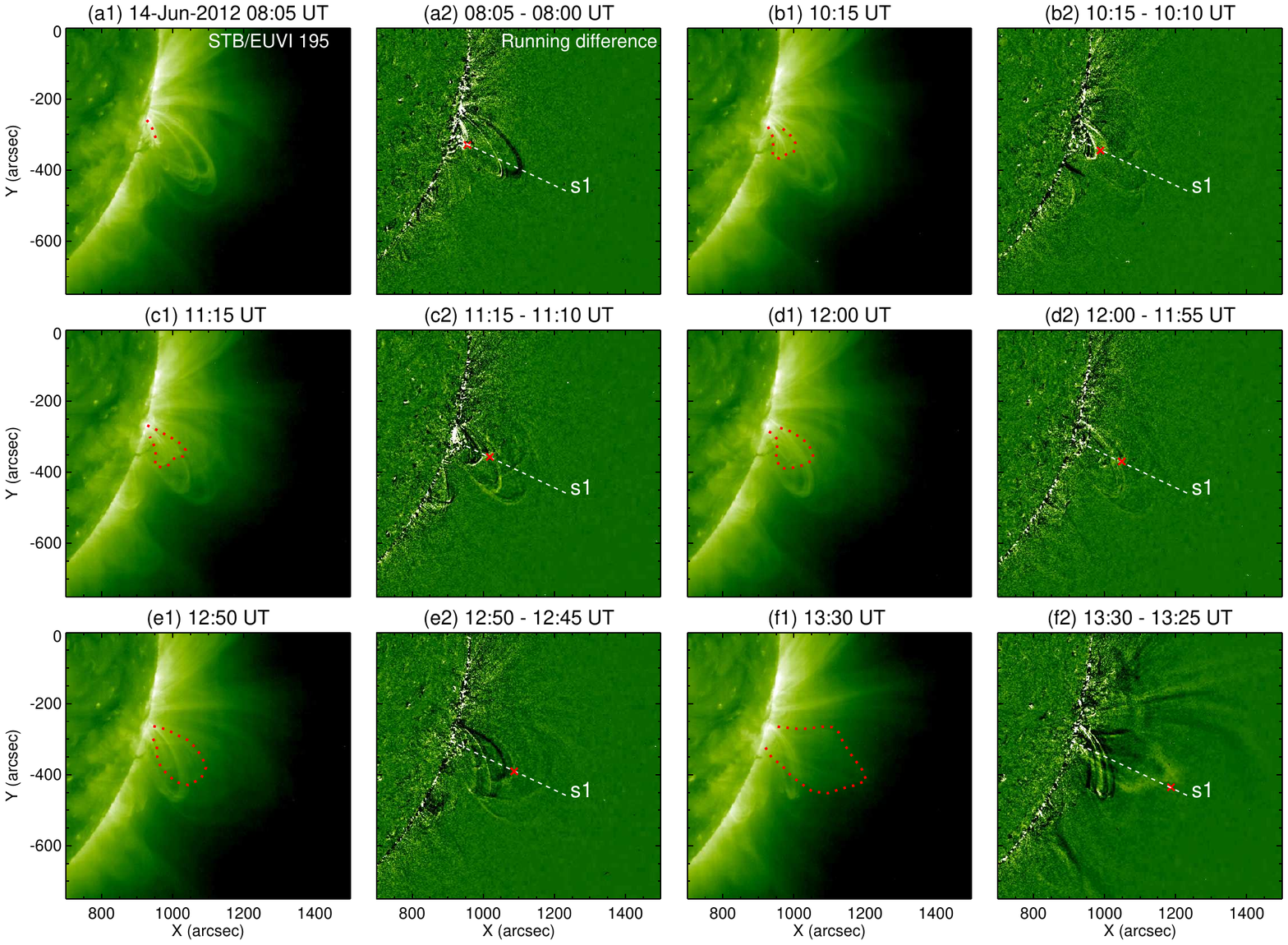}
\vskip -3cm
\caption{Snapshots of an expanding coronal structure identified in STB/EUVI 195 and the corresponding running-difference images. The structure is marked by a red dashed curve in each original image. Symbol `X' in each difference image indicates its position along the slit `s1' (Figure~\ref{fig4} (d)) at that moment. Also see the evolution of this structure in animation 1. \label{fig2}}
\end{figure}

\begin{figure}
\epsscale{.9}
\plotone{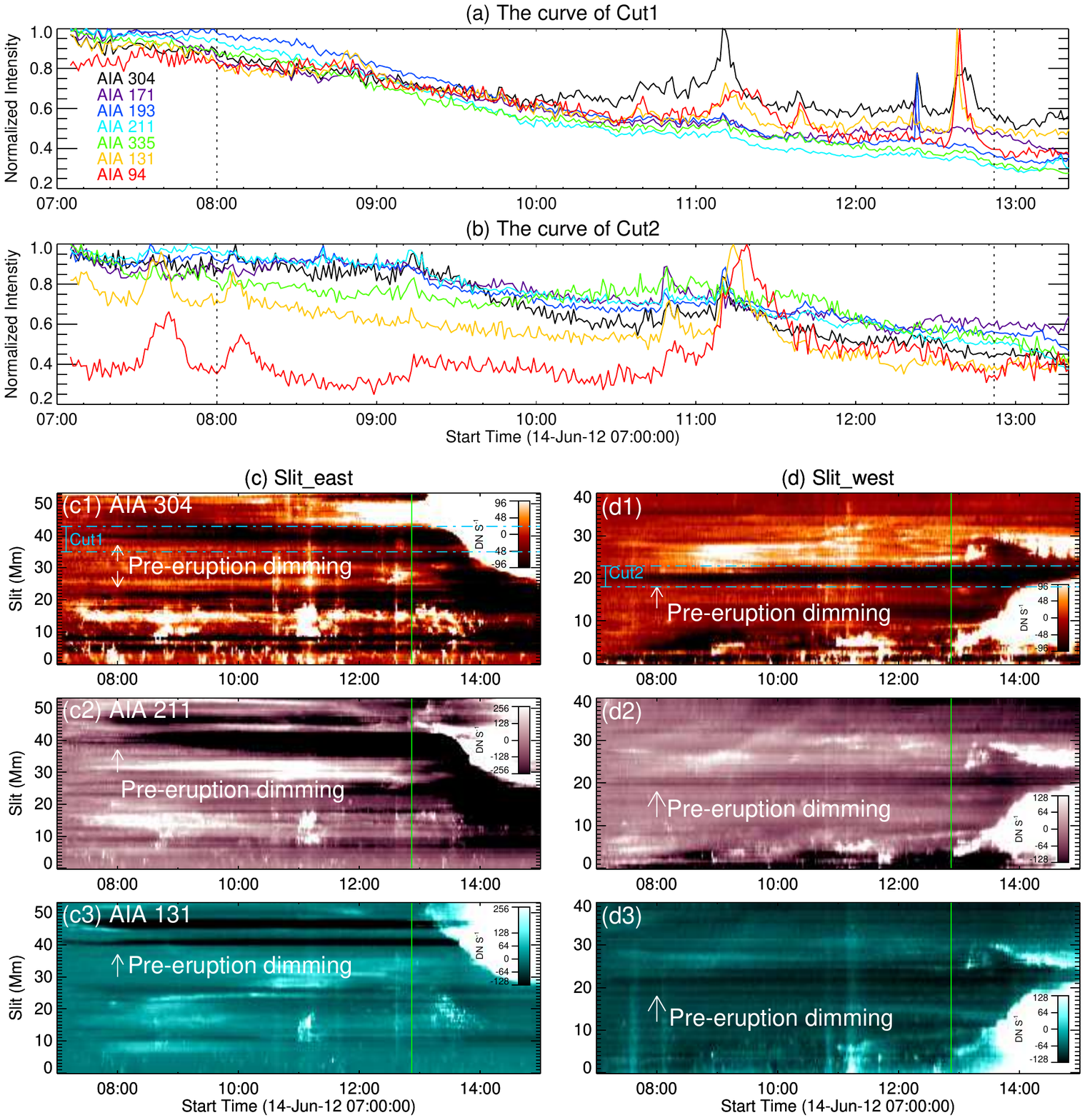}
\caption{Pre-eruption dimmings observed in seven AIA EUV channels. Panels (a) and (b) show the normalized lightcurves of the dimmed segments of Cut1 in (c1) and Cut2 in (d1) for seven AIA channels, respectively. Panels (c) and (d) at the bottom are time-distance maps along the two slits (Figure~\ref{fig1} (a)) from 304~{\AA}, 211~{\AA}, and 131~{\AA} base-difference images. ``Slit\_east'' starts from its western endpoint (c1 $-$ c3). ``Slit\_west'' starts from its eastern endpoint (d1--d3). The darker regions, marked by arrows, indicate the pre-eruption dimmings. Green vertical lines in panels represent the start time of the M1.9 flare. \label{fig3} }
\end{figure}

\begin{figure}
\epsscale{.90}
\plotone{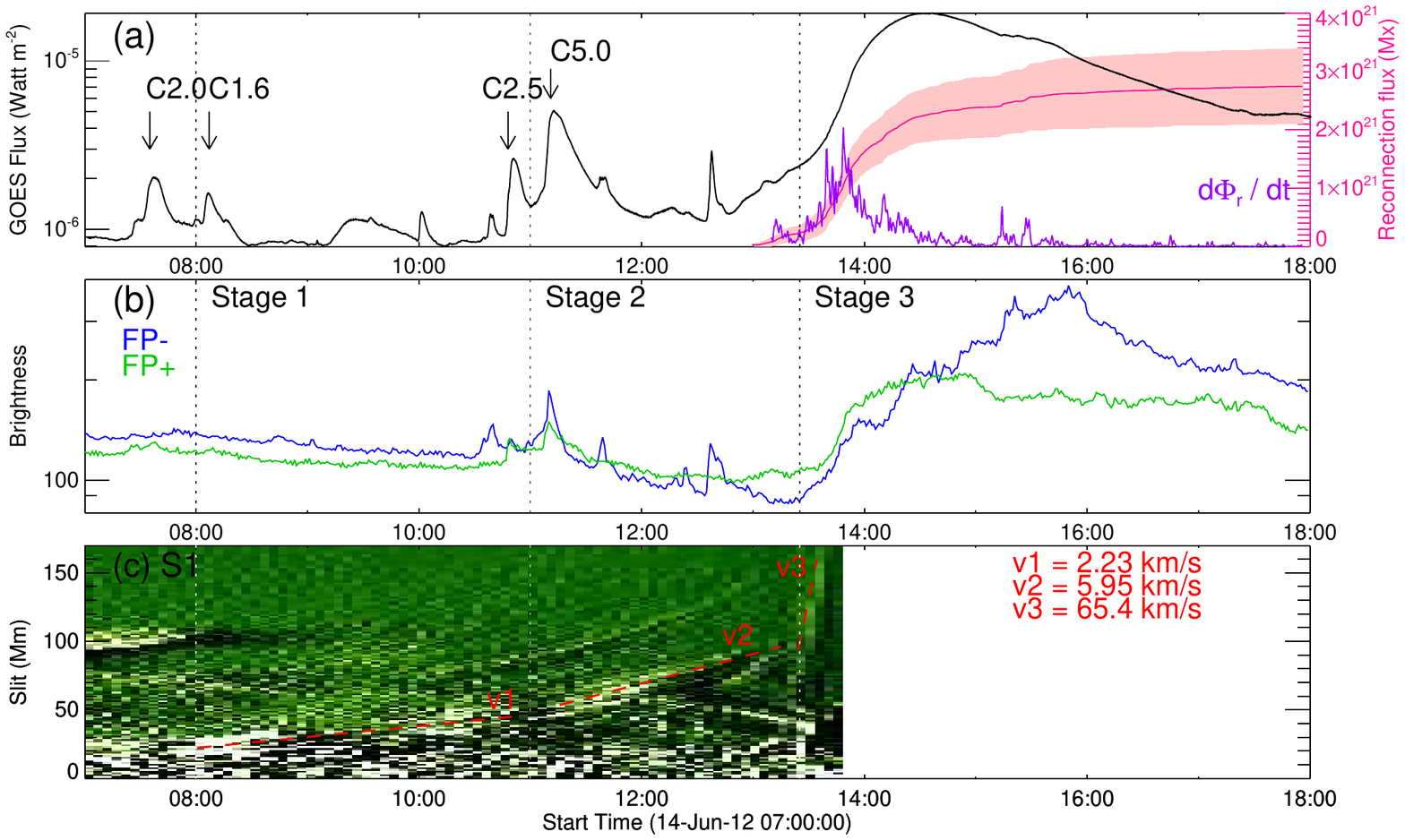}
\vskip -3cm
\caption{Evolution of the dynamic properties of the MFR during 07:00 UT to 18:00 on 2012 June 14. (a) The GOES 1$-$8~{\AA} light curve. Pink and purple lines represent the magnetic reconnection flux and its time derivative, respectively.  (b) The temporal profiles of brightness in AIA 304. Green and blue curves represent the average brightness in two fixed area (contours in Figure~\ref{fig1} (d)). (c) The time-distance map along `s1' in EUVI 195~\AA\ in Figure~\ref{fig3}.  Three linear fittings are indicated by red dashed lines. Three vertical dashed lines mark the beginning of three different stages. \label{fig4}}
\end{figure}

Dimming is usually related to reduced plasma density associated with an expanding coronal structure (\eg \citealt{harrison2000spectroscopic,harra2001material}). STB/EUVI 195~{\AA} images indeed reveal the slow rise of a coronal structure, which persists for five hours before the eruption. The evolution of this structure is outlined by red dashed lines in Figure~\ref{fig2} (a1)--(f1). To track its motion, a time-distance stack plot along a fixed slit of `s1' (Figure~\ref{fig2} (a2)) is generated. The red dashed lines in Figure~\ref{fig4} (c) indicate linear fittings of its trajectory along the slit at three stages coincident with those of the dimming evolution. It is notable that, from the first to the second stage, as the coronal structure expands more rapidly, the brightness in dimming regions also decrease more quickly. 

The expanding coronal structure finally erupts as a halo CME (Figures~\ref{fig1} (b) and (c)), which arrives the Earth two days later and is identified as an MC with strong magnetic fields peaking at $\sim$40 nT (\citealt{palmerio2017determining,james2017disc}). We employ the Grad-Shafranov reconstruction method \citep{hu2002reconstruction,qiu2007magnetic,hu2014structures} to derive the magnetic structure of the MC. The result suggests that it is a highly twisted flux rope with 4.0 turns per AU. In summary, the combined remote-sensing image and the in-situ detection support the existence of a pre-existing MFR, which roots in the conjugate dimming regions. \par

\subsection{Magnetic Properties}
We infer the magnetic properties of the MFR from its feet (FP+ and FP- in Figure~\ref{fig1} (d)). Table~\ref{tab:feet} lists the measurements at the two feet, including the mean vertical magnetic field $\langle B_z\rangle$, mean transverse field $\langle B_t\rangle$, total magnetic flux $\Phi_z$, mean vertical current density $\langle J_z\rangle$, and total current $I$. The FP+ is inside the leading sunspot with  $\langle B_z\rangle \approx$ 1500~G, magnetic flux $\Phi_+ \approx 4.2 \times 10^{21}$~Mx. In FP+, the positive current ($I_{+} \approx 1.7\times 10^{12}$~A) dominates over the negative current ($I_{-} \approx -0.5\times 10^{12}$~A), resulting in a net positive current $I_{net} \approx 1.3\times 10^{12}$~A. The FP- is located within a so-called magnetic tongue \citep{luoni2011twisted} near the polarity inversion line (PIL), with $\langle B_z\rangle \approx -700$~G, and $\Phi_- \approx -3.0 \times 10^{21}$~Mx. In FP-, the negative current ($I_{-} \approx -3.8\times 10^{12}$~A) is about twice the positive current ($I_{+} \approx 2.0\times 10^{12}$~A). As a result, FP- has a net negative current $I_{net} \approx -1.8\times 10^{12}$~A. Thus both $\Phi_z$ and $I$ at the two feet are comparable and of opposite signs, agreeing with the scenario that the conjugate dimming regions are connected by a current-carrying MFR. \par

\begin{figure}
\epsscale{.95}
\plotone{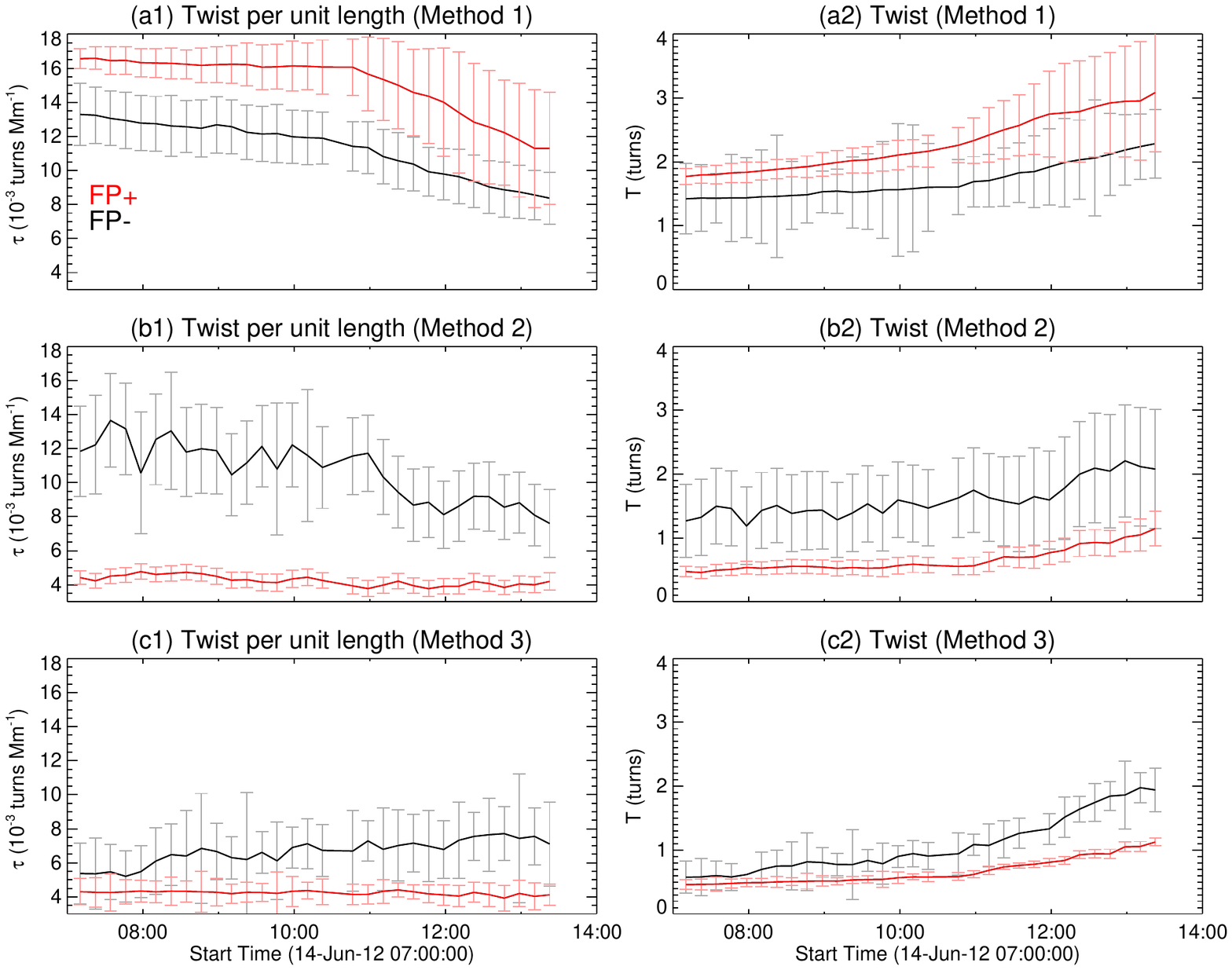}
\vskip -3cm
\caption{ Evolution of the magnetic twist during 2012-06-14 07:00 to 14:00 UT. Panels on the left shows twist per unit length, while panels on the right give the total twist. From top to bottom are the results from three methods. Red (black) curves represent the values calculated for FP+ (FP-). \bluec{Errors here come from two sources: the uncertainty of calculation is estimated using errors proprogation; and uncertainty of measurement that related to the pixel selection ($B_{z} >20$ G, $B_{t} > 100$ G).} \label{fig5} }
\end{figure}

Based on the observational measurements listed in table~\ref{tab:feet}, the magnetic twist $T$ of the MFR can be estimated with three methods. First, under a simplified assumption of an axial-symmetric cylindrical flux rope, the twist is given by \[T=\frac{LB_{\theta}(r)}{2\pi r B_{z}(r)} \qquad \qquad (1)\] in cylindrical coordinates $(r,\theta,z)$, where $L$ is the length of the MFR axis, and $r$ is the distance to the axis. The twist per unit length at each foot, $\tau = T/L$, is calculated utilizing the photospheric magnetogram. We treat the geometric center of each foot as the axis of the MFR; $r$ is the distance of a pixel to the center (or axis), and $B_{\theta}$ is calculated from the transverse field $B_t$. We hence obtain $\tau $ for every pixel in FP+ and FP-. We then shift the center by up to a quarter of the foot size, and find the uncertainty is about $1.2 \times 10^{-3} \ \text{turns\ Mm}^{-1}$. Then the total twist along the MFR is evaluated with the averaged value $\langle \tau \rangle$ by $T = \langle \tau \rangle L$. $L$ is determined by assuming a circular-arc shape of the MFR, utilizing the height of the coronal structure and the distance between two geometric centers of its feet. We derived the  errors of $B_{z}$ and $B_{t}$ from the uncertianties of HMI data \citep{liu2012comparison,hoeksema2014helioseismic} and found the errors vary within 20 Gauss for $B_{z}$ and 100 Gauss for $B_{t}$ during five-hour evolution. We hence calculate the twist using the effective pixels that satisfy two conditions ($B_{t} > 100$ G and $B_{z} > 20$ G). The time variation of $\langle \tau \rangle$ and $T$ in each region is plotted in Figure~\ref{fig5} (a). \par
 
The second and third methods both assume non-linear force-free magnetic configuration \citep{liu2016structure}. The twist of a field line about its own axis (rather than the axis of an imaginary cylinder) is given by \citep{berger2006writhe} \[T=\int_{0}^L \ \frac{  \mathrm{\mu_{0}}\mathrm{J_{\parallel}}}{ \mathrm{4} \pi \mathrm{B}} \ \mathrm{d}l \qquad \qquad (2)\] where $J_{\parallel}$ is the current density parallel to the magnetic field. Under the non-linear force-free assumption, $J_{\parallel}/B$ is constant along the same field line and is equal to $J_z/B_z$ at its feet, though varies across different field lines. Thus equation (2) gives $\tau = \ (\mathrm{\mu_{0}} \mathrm{J_{z}}) / (\mathrm{4} \pi \mathrm{B_{z}})$. We take the average value $\langle \tau \rangle$ over all pixels in FP+ and FP-, respectively. Figure~\ref{fig5} (b) shows the evolution of $\langle \tau \rangle$ and $T$. The third method is similar to the second one; instead of measuring $\tau$ at each pixel and taking the average over all pixels, we measure \[ T = \frac{\mathrm{\mu_{0}} \mathrm{I}}{ \mathrm{4} \pi \Phi} \mathrm{L} \qquad \qquad (3)\] where $I$ is the total current and $\Phi$ is the total magnetic flux in each foot. The results from this method are plotted in Figure 5 (c). \par

Without knowing the exact configuration of the MFR, we regard that the measurements with the above three methods provide a possible range of twist in the MFR. Figure 5 shows that the twists estimated by these methods exhibit similar behavior in general. Based on the three methods, the original value of $T$ is $1.7\pm0.4$, $0.9\pm0.5$, or $0.6\pm0.3$ turns, respectively, and it then increases to $2.8\pm0.4$, $1.6\pm0.6$, or $1.6\pm0.2$ turns correspondingly before the eruption. Despite discrepancies in exact values, all three methods suggest that the total twist $T$ of the MFR has increased by about 1 turn during the pre-eruption phase. 

The magnetic helicity inside the MFR can be estimated by $H = T\Phi^2$ \citep{berger1984topological,webb2010alfven}, where $T$ is the total twist and $\Phi$ is toroidal flux calculated by the effective pixels. The $\langle H \rangle$, averaged over the measurements, gradually increases from $6.0 \pm 3.8$ to $10.4 \pm 2.0$ ($\times ~ 10^{42}$~Mx$^{2}$) during the pre-eruption phase. To make a comparsion, we also investigate the relative helicity of the host active region. We calculated the helicity injection across the photospheric boundary $S$ of the active region using the following formula: \[ \frac{dH}{dt}|_{s} = 2 \int_{S} \ (\mathbf{A_{p}} \cdot \mathbf{B_{t}}) V_{\bot n} \ dS - 2 \int_{S} \ (\mathbf{A_{p}} \cdot \mathbf{V_{\bot t}}) B_{n} \ dS, \qquad (4)\] where $\mathbf{A_{p}}$ is the vector potential of the reference potential field; t and n refer to the tangential and normal directions, respectively; $\mathbf{V_{\bot}}$ is the photospheric velocity that is perpendicular to magnetic field lines, and is composed of a emergence term ($\mathbf{V_{\bot n}}$) and a shear term ($\mathbf{V_{\bot t}}$). Readers are referred to \cite{liu2016structure} and references therein for more information. The accumulated helicity calculated at the two feet of the MFR is about $1.2 \pm 0.4 \ (10^{42}$~Mx$^{2})$ during the five-hour evolution, which is translated to a twist number of about 0.2 turn. Also, the sunspot rotation could contribute a twist number up to 0.1 turn during this period.


\subsection{Relationships between the flares and the MFR evolution}
\begin{figure}
\epsscale{.95}
\plotone{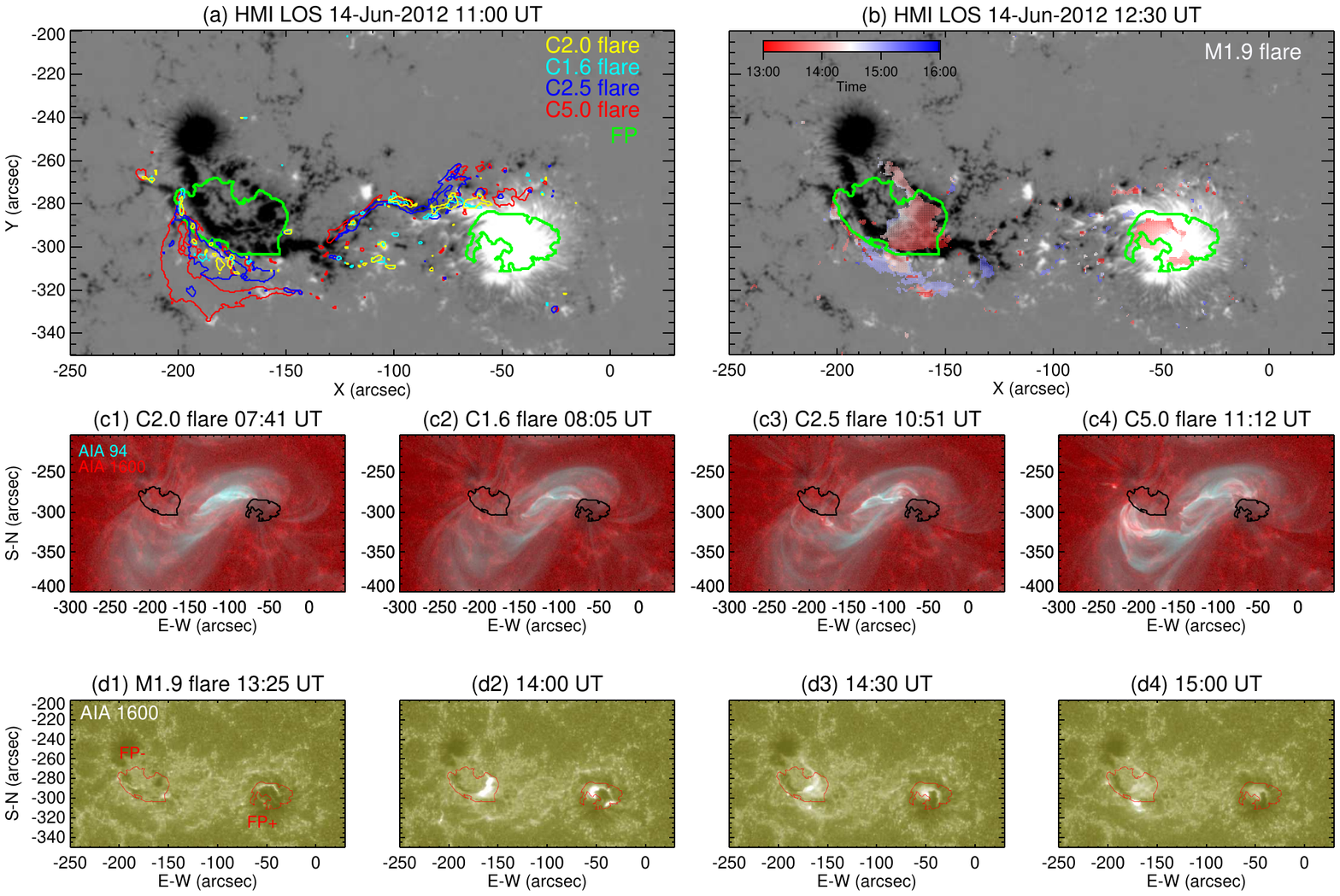}
\vskip -3cm
\caption{ The morphologies and motions of the flare ribbons. Panels (a) and (b) are two snapshots of AR~11504 in the HMI magnetograms along the line-of-sight. Panel (a) shows the ribbon locations of four C-class flares, with each flare indicated by a distinct color. Green contours mark FP+ and FP-. The sweeping motions of an M1.9 flare ribbons are given in (b). Panels (c1) to (c4) show snapshots of the four C-class flares in SDO/AIA 94~{\AA} (cyan) and 1600~{\AA} (red). Panels (d1) to (d2) give snapshots of the M1.9 flare in SDO/AIA 1600~{\AA}, showing that the ribbons sweep partially through FP+ and FP-. Evolution of these flares can be found in animation 2. \label{fig6} }
\end{figure}

The pre-existing MFR undergoes three stages, indicated by its rising motion at three different speeds and the evolution of the conjugate dimmings. It is noteworthy that each stage is preceded by flare(s) (Figure~\ref{fig4} (a)). At around 08:00~UT, two homologous C-class flares (C2.0 and C1.6, see Figures~\ref{fig6} (c1) and (c2)) occur successively. Meanwhile, the MFR begins to rise, and the dimmings start to develop at its feet. The slow expansion of the MFR continues until another set of homologous C-class flares (C2.5 and C5.0, see Figures~\ref{fig6} (c3) and (c4)) take place. After that, the MFR rises more rapidly, and the conjugate dimmings develop more steeply. The ribbons of these four flares (contours in Figure~\ref{fig6} (a)) are primarily located along the edges of the MFR's two feet. And the magnetic polarity of each ribbon is opposite to its nearby feet of the MFR, forming a quadrupole configuration. \par

The M1.9 flare occurs at around 12:50 UT. Different from the previous C-class flares, the ribbons of the M1.9 flare spread into the MFR's two feet, as seen in Figures~\ref{fig6} (d1) to (d4). Figure~\ref{fig6} (b) shows the movement of the M1.9 flare ribbons, with a different color at a varying time. At the maximum area, the flare ribbons have encompassed almost half the areas of MFR's feet. We measure the magnetic fluxes in the areas swept up by the newly brightened ribbons. These fluxes are equivalent to magnetic reconnection flux in the corona (\citealt{qiu2007magnetic}). Figure~\ref{fig4} (a) shows the reconnection flux and its time derivative, or the magnetic reconnection rate. It is noted that the reconnection rate grows abruptly at 13:25~UT, when the MFR is suddenly accelerated (Figure~\ref{fig4}~(c)) and then ejected. \par

\section{Discussion \& Conclusion} \label{dis}
The MFR studied in this work undergoes a complex three-stage evolution. It is of interest to determine the physical mechanism(s) behind the MFR evolution. In the following sections, we first discuss the possible initiation mechanisms, including ideal MHD instabilities and magnetic reconnection. We then focus on evolution of the host active region to discuss the formation of the MFR. \par

\subsection{Ideal MHD instabilities}
\begin{figure}
\epsscale{.9}
\plotone{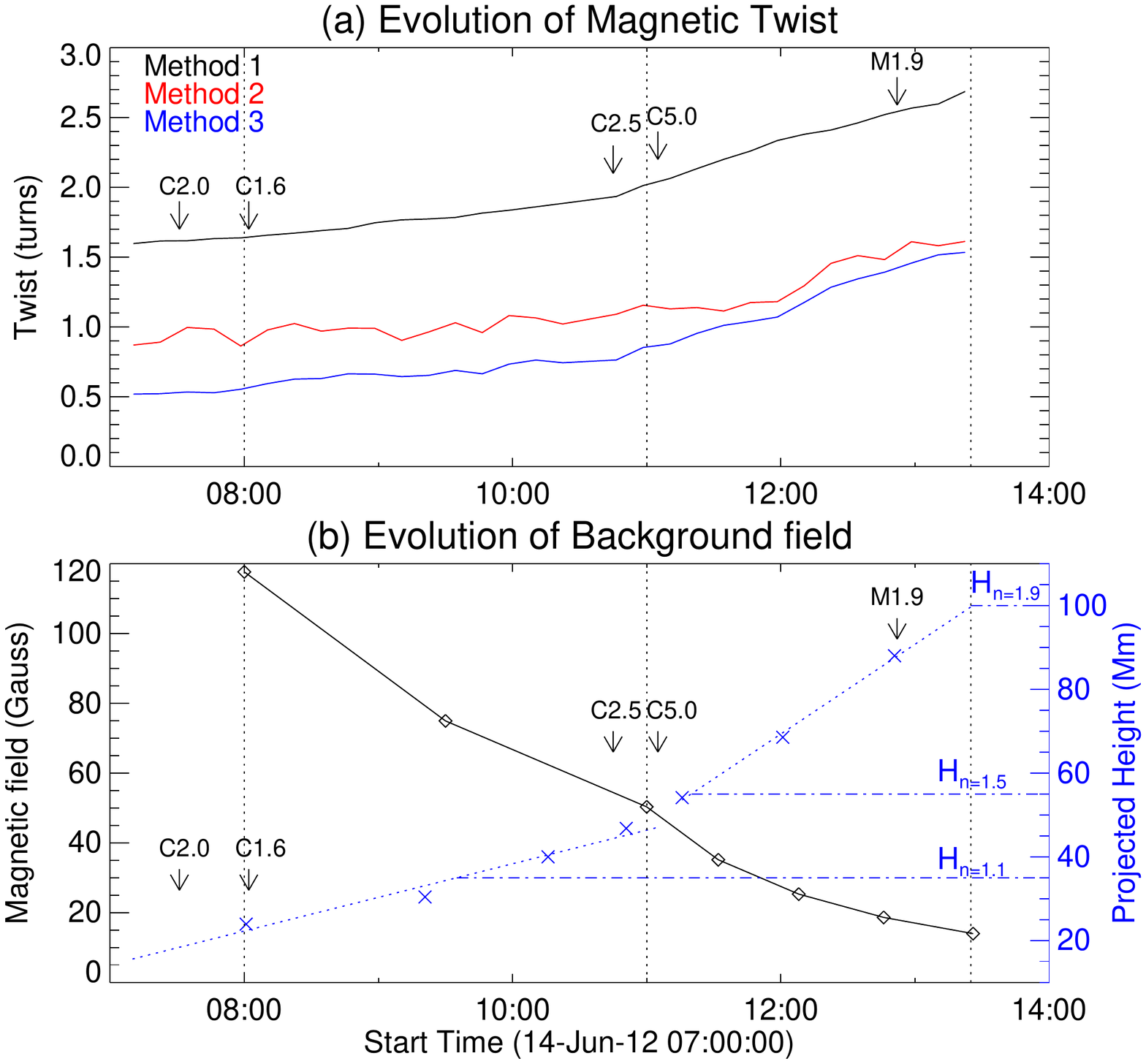}
\caption{Evolution of key parameters before the eruption. (a) Evolution of the twist from three methods. (b) The background field at the height of the expanding coronal structure. The black curve represents the variation of the field strength. Blue dashed line gives the height of the coronal structure. Vertical dashed lines in two panels label the three-stage evolution. Arrows mark the onset of flares. Blue horizontal lines in (b) show the decay indexes at corresponding heights.   \label{fig7}}
\end{figure}

Observations of this event provide a unique opportunity to examine the role of ideal MHD instabilities during the evolution of the MFR, such as kink and torus instabilities. In Figure~\ref{fig7}, we show the total twist of the MFR measured with three methods.  During these five hours, the MFR can be considered to be evolving in quasi-equilibrium. Theoretical and numerical studies have estimated $T_{crit}$ for kink instability to be somewhere between 1 to 2 turns, depending on the specific configuration of MFRs (\eg \citealt{dungey1954twisted,hood1979kink,bennett1999waves,baty2001mhd}). The quasi-equilibrium evolutions suggest that the kink instability might not occur during the pre-eruption phase, though the twist of the MFR may have exceeded $T_{crit}$ from some previous studies. Our measurements provide the value of $T_{crit}$ as 2.0 $\pm$ 0.5 turns, if the final eruption is triggered by kink instability. \par


To examine the possible role of the torus instability, we measure the decay index $n$ of the envelope magnetic field $B_p$ at the height of the MFR as it rises. We use the PFSS package in SolarSoftware, which takes into account the full sphere by assimilating magnetograms into a flux dispersal model \citep{schrijver2003photospheric}, and yields the coronal fields with a potential-field source-surface model at 12:04 UT.  Figure~\ref{fig7} (b) displays $B_p$ above the PIL at the CME's varying heights. As the MFR rises from 25 to 50 Mm in the first stage, $B_p$ decreases from 100 to 50~G, and the decay index increases from 1.0 to 1.5. In the next two hours, the MFR rises to 100 Mm at the onset of the eruption, while $B_p$ keeps decreasing to nearly 20~G, and the decay index has reached 1.9. Therefore, if torus instability triggers the MFR eruption, the decay index threshold would be $\sim$1.9 in this study, corresponding to the critical values of 1.1 -- 2.0 from previous works (\eg \citealt{fan2007onset,demoulin2010criteria,zuccarello2016apparent}).  \par

\subsection{Roles of the reconnection}
The observations of a series of flares indicate that reconnection may play important roles in the MFR evolution. The four C-class flares, with the ribbons located adjacent to the MFR's feet, probably help to reduce the constraints from the envelope fields, resulting in the continuous rising of the MFR. On the other hand, our measurements suggest the twist of the MFR increases by about 1.0 turn during its quasi-static evolution. The photosphere evolution can contribute to a twist number of 0.2 turn into the MFR, implying that most of increased twist may come from reconnection in the corona. This is reminiscent of numerical simulations by \cite{fan2010eruption}, which demonstrate that tether-cutting reconnections can add twisted flux to the MFR, allowing it to rise quasi-statically to the critical height of torus instability. We note that \cite{james2017disc} studied this event and concluded the MFR formed via tether-cutting reconnection, two hours before the eruption. Tether-cutting reconnection can indeed  convert sheared flux into the MFR. However, our analysis of SDO and STEREO observations suggest that the MFR, or at least its seed, was present at least five hours before the eruption, considering the synchronous evolution of the conjugate dimmings and the coronal structure, as well as the strong currents at its two feet.


The rapid acceleration of the MFR occurs in the impulsive phase of the M1.9 flare, similar to previous studies \citep{zhang2001temporal,gallagher2003rapid,cheng2003flux,qiu2004magnetic,zhang2006statistical,gopalswamy2012properties}. Noteworthily, the geometry of the M1.9 is quite different from the earlier C-class flares. The flare ribbons, which map the footpoints of newly formed loops via reconnection, spread into the two feet of the MFR, indicating that reconnection takes place between the two legs of the MFR (\eg leg-leg reconnection,  \citealt{kliem2010reconnection}).

\subsection{Evolution of the host active region}
Strongly non-neutralized currents in an active region may be an essential condition for eruptions. Recently, \cite{liu2017electric} found that the production of CMEs is related to the neutralization of the electric current in the host active region. They calculated the ratio of direct and return currents, and found it ranged from 2.0 to 3.0 in two active regions with CMEs, while it was about 1.0 for the other two active regions without CMEs. For this event, however, the ratio is about 1.0 in this active region with the CME, but at both feet of the MFR is about 2.9 (3.8 for FP+, 1.9 for FP-), suggesting the MFR carrying strong current may be essential for the eruption. To prove that, statistic studies are required in the future. \par

\begin{figure}
\epsscale{.9}
\plotone{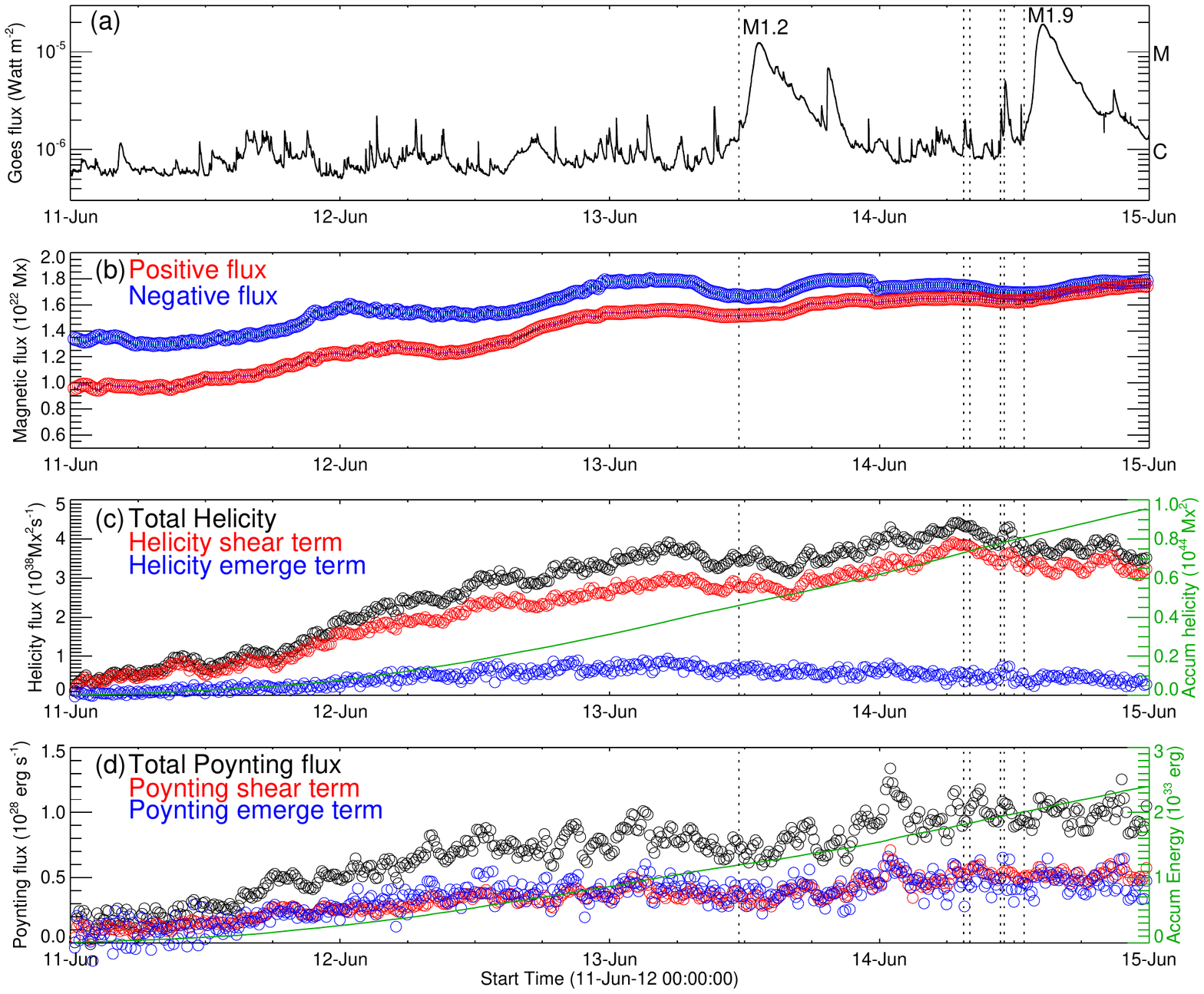}
\caption{Long-time evolution of AR~11504 from June 11 to 15. The panels from top to bottom show the GOES 1$-$8~{\AA} light curve, the magnetic fluxes, the helicity injection rate, and Poynting fluxes. The accumulated helicity and magnetic energy are displayed in green in the corresponding panels, scaled by the right y-axis. Six vertical dashed lines in all panels indicate the start times of the flares, including two eruptive flares (M1.2 and M1.9) and four confined flares (C2.0, C1.6, C2.5, and C5.0). \label{fig8}}
\end{figure}

Meanwhile, the formation of an MFR in the corona is thought to be associated with evolution of its host active region, \ie the injections of magnetic energy and helicity \citep{low1994magnetohydrodynamic,liu2010gradual}. The MFR in this study is found to carry a strong electric current rooted in the rotating sunspot and the magnetic tongue. Thus the long evolution of AR~11504 from June 11 to 15 are investigated, including the magnetic flux, the relative helicity and the Poynting flux, as shown in Figure~\ref{fig8}. The Poynting flux is calculated across the photospheric boundary $S$ of the active region using the following formula (see detail in Section 2.2 and Appendix B in \cite{liu2016structure}) \[ \frac{dP}{dt}|_{s} = \frac{1}{4\pi} \int_{S} \ {B_{t}}^2 V_{\bot n} \ dS - \frac{1}{4\pi} \int_{S} \ (\mathbf{B_{t}} \cdot \mathbf{V_{\bot t}}) B_{n} \ dS. \qquad (5) \] The total flux of the active region grows till mid-day June 13. The total helicity and Poynting fluxes into the active region keep rising, characterized by the sunspot rotation and the shear flows in the trailing plage. During the pre-eruptive evolution, the photospheric evolution can inject a twist number of 0.2 turn into the MFR. Note that the calculation of the helicity injection and the twist of the MFR follow the different assumptions. It is difficult to conclude the role of the photospheric evolution plays in the build-up of the MFR only from the quantitative difference. \par


\medskip
\subsection{Conclusions}
In this study, we analyze the evolution of the pre-eruption dimmings and an associated expanding coronal structure during five hours before its eruption into a halo CME, utilizing the triple-viewpoint observations from SDO and STEREO. Our study suggests the existence of a pre-existing MFR, with its two feet being mapped by the conjugate dimmings. The MFR undergoes a three-stage evolution i.e., a two-stage gradual expansion followed by another rapid acceleration/eruption. Moreover, quantitative measurements indicate that magnetic twist of the MFR increases by about 1 turn during the two-stage gradual expansion. \par

The results raise two major questions for this study: what causes the MFR's gradual expansion? how does magnetic twist increase during the expansion of the MFR? As discussed in previous sections, the four confined flares might help weaken the envelope field of the MFR, resulting in the expansion. On the other hand, magnetic reconnection may play the major role in the increase of the MFR's twist during the five-hour evolution. Another important question is what triggers the MFR eruption. If ideal MHD instability is responsible for the eruption, our study gives the critical value of 2.0 $\pm$ 0.5 turns for kink instability and the critical decay index of 1.9 for torus instability. The MFR erupts in the impulsive phase of the M1.9 flare, implying reconnection may also trigger the eruption. During the eruption, the leg-leg reconnection may have contributed a significant amount of twist into the MFR, such that the MC possesses a highly twist of 4.0 turns per AU, comparing with other MCs in a statistic study \citep{hu2014structures}.  \par

This study raises many other interesting questions. For example, how often can pre-eruption dimmings be observed? What are the temporal and spatial relationships between ribbons and dimmings if both observed? How often can MFRs be found to carry strong currents? To answer these questions, further investigations are underway. \par

\acknowledgments
We thank Dana W. Longcope for helpful comments. W.W. acknowledges the support from the China Scholarship Council (CSC) under file No.~201706340133. R.L. acknowledges the support by NSFC 41474151, 41774150, 41761134088. Q.H. acknowledges the support by NSF AGS 1650854. 
\\\textit{Facilities}: SDO, STEREO, GOES



\bibliographystyle{aasjournal} 
\bibliography{References}

\begin{deluxetable*}{CCRRRRRRR}[b!]
\tablecaption{Magnetic properties at the conjugate feet of the MFR \label{tab:feet}}
\tablecolumns{8}
\tablenum{1}
\tablewidth{0pt}
\tablehead{
\colhead{} & \colhead{} & \colhead{$\langle B_{z} \rangle$} & \colhead{$\langle |B_{t}| \rangle$} & \colhead{$\Phi_{z}$} & \colhead{$\langle J_{z} \rangle$} & \colhead{$I$} \\
\colhead{} & \colhead{} & \colhead{(G)} & \colhead{(G)} & \colhead{($10^{20}$ Mx)} & \colhead{(mA m$^{-2}$)} & \colhead{($10^{12}$ A)}
}
\startdata
 FP_{+} & (+) & 1555 \pm 35 &             & 41.75 \pm 0.90  &  9.2 $\pm$ 0.5   & 1.70 \pm 0.05   \\ 
        & (-) & 0           &             & 0.00           & -5.5 $\pm$ 0.4   & -0.45 \pm 0.03    \\
        &  \text{net}   & 1555 \pm 35 & 1084 \pm 48 & 41.75 \pm 0.90  &  4.7 $\pm$ 0.2   & 1.26 \pm 0.05    \\
\hline
 FP_{-} & (+) & 162 \pm 46  &             &  0.12 \pm 0.03  &  13.1 $\pm$ 0.8  & 1.98 \pm 0.06   \\ 
        & (-) & -734 \pm 37 &             & -30.62 \pm 1.00  & -16.4 $\pm$ 1.2  & -3.75 \pm 0.10   \\
        &  \text{net}   & -710 \pm 45 & 508 \pm 70  & -30.50 \pm 1.00  & -4.0 $\pm$ 1.0  &  -1.77 \pm 0.11 \\
\enddata
\tablecomments{ Table~\ref{tab:feet} shows the physical values related to the magnetic properties at the conjugated feet of the MFR (see contours in Figures~\ref{fig1} (d) and (e)). The subscript `z' represents the direction along the vertical, while `t' denotes the horizontal direction. The `+' (`-') rows give values from the pixels of a positive (negative) polarity. Errors of magnetic field ($\delta_{B}$) come from uncertainties in the HMI data. Then we used Monte Carlo simulation to provide uncertainties of current density ($\delta_{j}$) using $\delta_{B}$. The uncertainties of magnetic flux ($\delta_{\Phi}$) and current ($\delta_{I}$) are calculated from formulas of errors propagation.}
\end{deluxetable*}

\end{document}